\documentclass{article}
\usepackage{FPSpro}
\begin{document}
\title{ 
THE SIMULATION OF ASTROPARTICLE EXPERIMENTS\\AND THE GEANT4 TOOLKIT
  }
\author{
  Alessandro De Angelis        \\
  {\em Dipartimento di Fisica dell'Universit\`a di Udine and INFN Trieste} \\
  }
\maketitle
\baselineskip=11.6pt
\begin{abstract}
  The new generation of astroparticle experiments will use technologies 
common to
High Energy Physics. Among such technologies a central place is given
to the role of simulation.
\end{abstract}
\baselineskip=14pt
\section{Introduction}

Monte Carlo simulation is nowadays an essential tool in the project of experiments. It is so common for physicists that 
one could think that such a technique 
is in use since ages; but indeed it is quite young, 
around 50 years. It was suggested by Ulam and von Neumann\cite{von} in 1947
and
first used for particle transport by 
Wilson\cite{wilson}, in the context of the 
the problem of electromagnetic showers. 
Such a problem had an elegant solution by Rossi and Greisen\cite{rossi}, 
although based on approximations valid beyond 
energies of a few MeV.
 
To solve the integro-differential equations of showers in lead, 
Wilson in 1952 uses the following procedure.

{\em ``The procedure used was a simple graphical and mechanical one. 
The distance into the lead was broken into intervals of one-fifth 
of a radiation length (about 1 mm). The electrons or photons were followed through successive intervals and their fate in passing through a given interval was decided by spinning a wheel of chance; the fate being read from one of a family of curves drawn on a cylinder (...)
A word about the wheel of chance. The cylinder, 4 in. outside diameter by 12 in. long is driven by a high speed motor geared down by a ratio 20 to 1. The motor armature is heavier than the cylinder and determines where the cylinder stops. The motor was observed to stop at random and, in so far as the cylinder is concerned, its randomness is multiplied by the gear ratio (...)"}

What are the requirements of a general-purpose simulation software for particle transport? 
To start, one has
to take care in  an adequate way of the physics, i.e., of the interactions: both for what is related to the probability
of occurrence (i.e., to the cross section) and to the modelling of the final state. 
The physics of electromagnetic interactions is based on QED, and thus it is
in principle well known. Electromagnetic interactions should be 
well modelled down to low energies (how low depends
on the geometry and on the physics of the detector).
The theory of hadronic interactions, QCD, is not in a status comparable 
with QED;
as a consequence the simulations of hadronic interactions 
relies in general on QCD-inspired
models rather than on the theory itself,
and in general a reasonable simulation of hadronic interactions is 
good enough.

For a software to be used in the context of detector project and data 
analysis, however,
the physics requirements are not the full story, and maybe they are not even the most important part.
Technical requirements rely mostly on a well written code, with characteristics of
\begin{itemize} 
\item modularity;
\item easiness to add different generators;
\item easiness to add new physics routines;
\item friendly interfaces;
\item good documentation;
\item maintenability;
\item support on different platforms;
\item last but not least, a term taken from industry: customer care.
\end{itemize}

\section{The shop-list before Geant4}
Most of the large experiments
'90s were basing their simulation 
on the Geant3\cite{geant3} package. The Geant3 code comes from a long program
of development based at CERN between 1982 and 1994.
Geant3 is written in FORTRAN; among his strong points, compared to the other simulations, 
are user-friendly packages for geometry description and visualization,
and an overall easiness-of-use.
Geant3 has proprietary routines for electromagnetic physics,
and can bind several hadronic codes (GHEISHA is the most common).

However, a different package was and still is
the reference for electromagnetic physics: EGS\cite{egs},
a package which had a very long development and debugging at SLAC, LNL
and KEK 
in the period between 1966 and 1985. EGS, presently at the version 4 (EGS4),
is still the reference now for dosimetry, where
definetly one should not be wrong.
EGS4 has proprietary routines
for electromagnetic physics;
it is most commonly used in couple with FLUKA for hadronic interactions.
EGS4 is written in MORTRAN, a pre-processor of FORTRAN,
although several FORTRAN and C++ versions circulate on
the Web. The code is overall a bit unfriendly: the
geometry is difficult to define (it must be put in by means of a subroutine);
few facilities, in particular for visualization, are available.
The cross sections are computed by an offline preprocessor, PEGS,
which must be run separately.

Besides Geant3 and EGS4, very little room was available for other products
(DELSIM, GISMO etc.); and the success of a simulation code is boosted by the
widespread use (which guarantees updating, debugging, availability on different 
platforms, interfaces).

\section{The Geant4 toolkit}

At the end of the '90s most of the simulation programs in High Energy
Physics were based on Geant3. Several reasons however pushed the community to 
start a new project, aimed to improve Geant3. Among such reasons:
\begin{itemize}  
\item limitations of Geant3 maintenance; in particular, 
because of too complex structure driven by historical 
reasons, it was almost impossible to add a new feature or to hunt a bug;
\item limitation of FORTRAN, and choice of  object orientation and C++
by the physics community; 
\item shortage of man power at CERN;
\item limitation of ``central center'' supports. 
\end{itemize}
Such reasons convinced a group of physicists and computer scientists 
at CERN to
start a world-wide collaboration for a new simulation project. 
Such a project was based  on the most recent software engineering 
methodologies 
and, from an organizational point of view, on a world-wide collaboration.

\subsection{Geant4: philosophy, history, future}

Geant4  aims to be the successor of Geant3 by
re­designing a major package of CERN software for the 
next generation of HEP experiments using an Object Oriented philosophy.
The final aim of the project is to build a
simulation more precise than EGS, and more friendly than Geant3.

A variety of new requirements also came from heavy ion 
physics, CP violation physics, cosmic ray physics, 
medical applications and space science applications
In order to meet such requirements, a large degree of 
functionality and flexibility are provided: Geant4 is not only for 
High Energy Physics.

The main steps of the history of Geant4 are, for the time being:
\begin{itemize}
\item[Dec 94]  Project starts
\item[Apr 97]  First alpha release 
\item[Jul 98]  First beta release 
\item[Dec 98]  Release 0.0 
\item[Jun 01]  Release 3.2 (with a complete list of physics processes)
\end{itemize}

Maintainment and upgrade are expected for at least 10 years; 
development is continuous, with two major releases each year 
plus a monthly  internal tag (frequent bug fixes, new features, new examples).

Geant4 is presently based on more than 700,000 lines of code.

For more details on the collaboration and an updated status, see the 
very well done and maintained web pages\cite{geant4}.

\subsection{The physics of Geant4}

The library proposes several models for the most important interactions 
of particles with matter.

In particular for electromagnetic interactions 
one can use a ``standard'' package with at least all the features of 
Geant3, or call a package specialized in the low energy part (which aims 
to an accurate modelling of physics down to 250 eV and below, based on 
an important use of experimental tables). On the other direction Geant4 should
be able to reproduce interactions up to the PeV and beyond: 
Geant4 is developed by people involved in fields other than 
High Energy Physics (e.g. medical physics, astroparticle physics).

All processes are already at least at level of Geant3,
and in addition there are new processes
(transition radiation, optical processes) and substantial improvements have 
been done in particular on the 
multiple scattering (new model, without
path length restriction and with lateral displacement), on energy loss
and on hard processes: in the future also the 
photoproduction of hadronic resonances will be modelled.

The electromagnetic  processes are going through extensive tests,
comparing both
with data and with Geant3-based and EGS4-based simulations.
Very good agreement with the data is seen on the simulation of 
electromagnetic showers.

For what is related to hadronic interactions one can bind GHEISHA,
but more performant models have just been released for the Geant4 code
and are under test.
There is a large variety of models according to the energy, including
string models
(Geant4 is interfaced with Pythia7 for hard scattering),
cascade models, 
evaporation
and break-up.

In any case Geant4 is an open system to new inputs:
the framework is such that different models can be easily integrated.

\subsubsection{Confrontation with data}

Facing such a huge development effort, debugging and tests 
is a major enterprise.

Fortunately, the distributed organization helps 
in boosting the manpower. Many comparisons 
have been done, and results published;
a lot more are ongoing or starting
within the collaborations using Geant4, in particular
the  LHC collaborations (ATLAS, CMS, LHCb, ALICE),
BaBar (migrating from Geant3)
and GLAST. 

Again the most important results are summarized in the Geant4 web page.

\subsection{Miscellaneous features}

Geant4 includes several features goring in the direction of functionality 
and easiness of use; many such features come from comments originated from 
Geant3 users.

In particular such improvements affect the definition of cutoffs,
geometry and utilities,
hits and digitizations,
fast simulation,
visualization.

\subsubsection{Cutoffs}

Contrary to what done in Geant3,
cuts in Geant4 are done in range rather than energy.
It makes poor sense to use the energy cut­off:  for example
the range of a10 keV gamma in Si is a few cm, while  
the range of 10 keV electron is a few micron.

This modification causes a significant gain in results quality 
versus CPU usage; however, users can override the default and
impose a cut in energy, track length, or time-of-flight.
Physics processes can also   ask to override the default 
when they need to (for example for a better treatment of boundary effects).

\subsubsection{Geometry}

Like in the philosophy of Geant3, Geant4 pre-defines 
basic geometries. The user can build 
new solids from union, intersection, subtraction 
of two solids (boolean solids) plus a transformation.
 
A utility {\em g3tog4} is provided to convert a Geant3 geometry into Geant4.

An interface with XML is in progress.

\subsubsection{Hits and digits}
Each logical volume can have a pointer to a sensitive detector;
a hit is a snapshot of the physical interaction of a track or an 
accumulation of interactions of tracks in the sensitive detector.

A sensitive detector creates hit(s) using the information given in 
a tracking step; the user has to provide his/her own implementation 
of the detector response.
A digitization is created with one or more hits and/or 
other digits by an explicit implementation 
by the user.

\subsubsection{Fast simulation}
Geant4 allows to perform full simulation and fast simulation 
(based on shower parametrizations, less accurate cutoffs etc.)
in the same environment.
The fast simulation produces the same objects 
as the full simulation (tracks, clusters etc.)

The full design is such to guarantee flexibility:
the user can activate fast/full simulation by detector and/or by particle type,
and use parallel geometries.

\subsubsection{Visualization}

Geant4 provides interfaces to graphics drivers 
(DAWN, RayTracer, OPACS, OpenGL, OpenInventor, VRML)
such that one can visualize detector, hits and trajectories.

\subsection{Things one has to do to run Geant4}

Documentation 
(Getting started and installation guide,
User guide for application and toolkit developer
Software and physics reference manuals) is available at the Geant4 web site.

For many users, however, starting from the study of a 
manual is not the most effective way. For such users examples
are provided: they
can go to the Geant4 Web site, 
run an example and see how it is done.

Six novice examples are available
with simple detectors and different experiment types
to demonstrate the essential capabilities of Geant4:
transport of a non-interacting particle through a slab,
track in a simplified tracking detector,
electromagnetic shower (full),
particle collision,
parametrised electromagnetic shower,
optical photon.

In addition advanced examples are available, two of which
are relevant for astroparticle:
\begin{itemize}
\item xray\_telescope, illustrating an application 
for the study of the radiation background in a typical X-ray telescope;
\item gammaray\_telescope, illustrating a gamma satellite-based
 detector of the new generation, similar to AGILE and GLAST\cite{gammar}. 
\end{itemize}

\subsection{Experience with Geant4}
The production release is in use by many experiments in High Energy Physics and
Astroparticle and by groups involved in medical physics.

Thanks to the wide use, the Geant4 developers got feedback.
The first results confirm some of the Geant4 strengths
in performance, simplicity of use, electromagnetic physics. 

Benchmarks between Geant3 and Geant4 in electromagnetic showers 
demonstrate that
Geant4 gives better physics at the same speed
(and better speed for same physics).

The tests evidence also some weaknesses; reaction is fast.

\section{The simulation of GLAST}
An example of implementation for 
space applications is the simulation of the GLAST gamma-ray
 telescope\cite{glast}.

GLAST has a wide range of physics objectives, from
gamma astrophysics to fundamental physics. Correspondingly, the simulation 
should have an easy interface to the simulation of different sources, and
be adequate both for the design and the physics analysis.

In addition, the gamma simulation in 
the tracker and in the calorimeter needs different details,
and in particular a fast simulation should be available for the huge hadron 
background.

The GLAST simulation has been done, from the beginning, using C++ and with OO technologies in mind (GISMO was the choice, also because
no other candidate present at that moment apart from standard FORTRAN 
simulations).

\begin{figure}[t]
  \vspace{11.0cm}
  \begin{center}
  \includegraphics{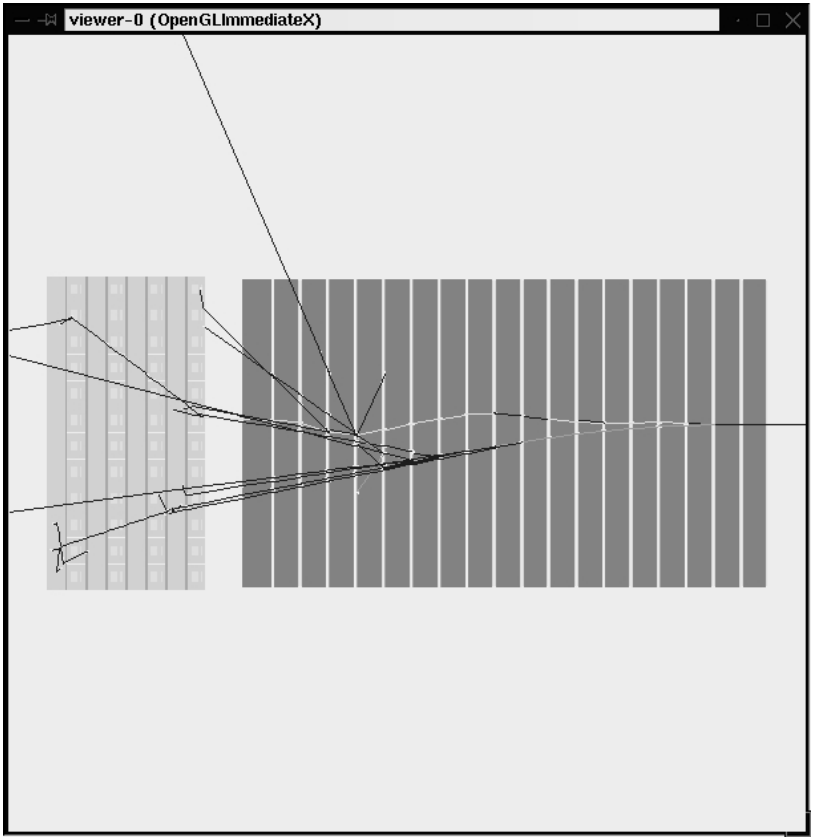} \end{center}
  \caption{\it
    Display of the interaction of a gamma ray with the detector in the
   Geant4 simulation of GLAST.
    \label{exfig} }
\end{figure}

The migration of the GLAST simulation to Geant4 is now almost complete;
it uses a prototype of the
XML interface for geometry description.

\section{Conclusions}
Geant4 has demonstrated to be suitable as a Monte Carlo toolkit, 
in particular for applications in astroparticle physics and High Energy
Physics. Among its strong points are the open structure 
(making it easy to integrate with specialized software)
and the easiness of use.

The communities of  astroparticle physics and High Energy
Physics are quickly acquiring a good experience, and the validation with 
data, formulae and standard simulations is progressing fast.

In conclusion, Geant4 is becoming the standard {\em de facto} both for 
the simulation of detectors and for particle and radiation transport.

\section*{Acknowledgements}
Many thanks to Riccardo Giannitrapani, Francesco Longo and Maria Grazia Pia 
for material and useful discussions.

\end{document}